\documentclass[fleqn,usenatbib]{mnras}

\usepackage{newtxtext,newtxmath}

\usepackage[T1]{fontenc}

\DeclareRobustCommand{\VAN}[3]{#2}
\let\VANthebibliography\thebibliography
\def\thebibliography{\DeclareRobustCommand{\VAN}[3]{##3}\VANthebibliography}

\newcommand{\lxo}{$L_{\mathrm{[OIII]}}$ --$L_{\mathrm{X}}$}

\usepackage{amsmath}
\usepackage{hyperref}

\usepackage{graphicx}	
\usepackage{amsmath}	

\title[VLT-MUSE spectroscopy of X-ray AGN]
{VLT-MUSE spectroscopy of AGNs misclassified by BPT diagnostic or with weak emission lines}

\author[Agostino et al.]{
Christopher J. Agostino, Samir Salim,$^{1}$ M\'ed\'eric Boquien$^{2}$, Steven Janowiecki$^{3}$, \newauthor H\'{e}ctor Salas$^{4}$, Guillherme S. Couto$^{4}$
\\
$^{1}$Indiana University, 727 East 3rd St. Swain West 318, Bloomington, IN 47405-7105, USA
\\
$^{2}$ Instituto de Alta Investigación, Universidad de Tarapacá, Casilla 7D, Arica, Chile\\
$^{3}$ University of Texas, Hobby-Eberly Telescope, McDonald Observatory, TX 79734, USA\\
$^{4}$ Leibniz-Institut für Astrophysik Potsdam (AIP), 
An der Sternwarte 16,
14482 Potsdam, Germany\\
}

\date{Accepted 02-Oct-2023.}

\pubyear{2015}

\begin{document}
\label{firstpage}
\pagerange{\pageref{firstpage}--\pageref{lastpage}}
\maketitle

\begin{abstract}
Despite powerful X-ray emission, some AGNs are known to either lack optical emission lines (so-called ``optically dull'' AGNs) or have lines that fall on the star-forming branch of the BPT diagram (``misclassified'' AGNs). Aperture effects have been proposed to explain such atypical spectra, especially when based on SDSS (3\arcsec) fibers.  We use observations from VLT-MUSE with Adaptive Optics to explore the spatially resolved optical emission line properties of 4 optically dull and 1 misclassified X-ray AGN candidates. VLT-MUSE IFU spectra allow us to investigate the extent to which the aperture size affects the emission line measurements. The optically dull AGNs become detectable in deeper VLT-MUSE spectroscopic apertures having the same size (3\arcsec) as SDSS fibers, suggesting no AGN is truly lineless. However, in no case does the line become more detectable as the aperture decreases, as would be expected if dilution by strong continuum was responsible for making the lines appear weak. We also show that the misclassified X-ray AGN retains the same position on the BPT diagram in smaller apertures (down to 0\farcs 5), demonstrating that its misclassification is not the result of the dilution by HII regions. Thus, we conclude that continuum swamping or star formation dilution, i.e., aperture effects, are not responsible for atypical lines. Rather, the AGN lines are intrinsically weak.

\end{abstract}



\section{Introduction} \label{sec:intro}

Accreting supermassive black holes---referred to as Active Galactic Nuclei (AGNs)--are believed to play a significant role in the cosmic evolution of galaxies through the process of feedback, wherein the radiation and jets produced by the accretion process may sufficiently alter the conditions of the ISM of the AGN host galaxy \citep{bower2006_feedback,cattaneo2009_feedback,harrison2017_feedback_review}. AGN feedback has been suggested as an explanation for various observed phenomena such as the quenching of star formation, but determining the extent to which the feedback plays a role in the host's evolution has been an elusive task, in part because of the difficulties and uncertainties in distinguishing which galaxies host AGNs versus those that do not. Thanks to surveys like the Sloan Digital Sky Survey (SDSS), the number of galaxies and AGNs which can be probed in statistical studies has increased dramatically, but the uncertainty surrounding the distinction of AGN hosts has been further exacerbated by possible host contributions on the AGN emission lines, necessarily included within the observations as a result of SDSS using a fixed aperture size (3\arcsec). 

Indeed, one of the primary areas where understanding the aperture effects is of critical importance is in the classification of galaxies based on their optical emission lines. Optical emission line diagnostic diagrams are used for separating AGNs and non-AGNs, with the Baldwin-Phillips-Terlevich \citep[BPT, ][]{bpt1981} diagram being the most widely used tool for doing so. Additionally, its completeness in identifying AGNs which have been identified using other AGN signatures, such as X-rays or infrared colors, has not been thoroughly verified.

 A variety of studies (e.g., \citealt{hornschemeier2005, georgantopoulos2005, pons2014, lamassa2019} and \citealt{agostino2019}) have demonstrated that there are X-ray-selected AGNs which either have weak emission lines or lines that do not appear to arise from AGN activity but predominantly from star formation. Host contributions within a spectroscopic aperture have been suggested to play a role in rendering the lines undetectable, but the extent to which this can account for the anomalously weak lines in some X-ray AGNs has not been established, and so there is considerable uncertainty. Thus, an accurate interpretation of AGN emission lines in SDSS rests on an understanding of the effect of including host contributions in the 3\arcsec\ aperture.  Furthermore, several studies (e.g., \citealt{yan2011, pons2014} and \citealt{agostino2023}) have used X-ray AGNs to assess the reliability and the completeness of the BPT diagram, finding that some X-ray AGNs cannot be reliably classified with the BPT diagram because of low S/N in one or more of the emission lines required for the full BPT classification. Such X-ray selected AGNs with weak optical emission lines have been called `optically dull' or X-ray Bright Optically Normal Galaxies (XBONGs; e.g. \citealt{comastri2002, trump2009,smith2014}) and have been suggested to lack optical emission lines for a variety of reasons: lines may be diluted by strong continuum light within the spectroscopic aperture \citep{moran2002}, a torus of dust with a high covering factor is blocking the light along our line of sight \citep{barger2001, comastri2002, civano2007}, the dust of the host is sufficient to attenuate the emission lines \citep{rigby2006}, a radiatively inefficient accretion flow (RIAF) cannot heat the narrow-line region (NLR) sufficiently so as to produce the emission lines \citep{yuan2004, hopkins2009, trump2009, trump2011a, trump2011b}, or the geometry and distribution of gas in the narrow-line region is such that it does not absorb enough photons to produce the emission lines of typical AGNs.
 Following similar ideas put forward in \citet{moran1996, barger2001, maiolino2003}, \citet{agostino2019}, argued that some X-ray AGNs are not detectable in SDSS spectroscopy because their lines are genuinely too weak and that probing them with higher spatial resolution spectroscopy will likely not reveal strong AGNs being drowned by their host contributions. 


This paper is organized as follows: in Section \ref{sec:data_methods}, we describe the sources of data, the selection of our parent and target samples, and the reduction of IFU data obtained with VLT-MUSE; in Section \ref{sec:results}, we describe our results; in Section \ref{sec:discussion}, we discuss the implications of our results in the context of previous studies; finally, in Section \ref{sec:conclusions}, we summarize our results. Throughout this work, cosmological parameters from \citet{planck2016} are assumed. 
\section{Data and Sample}  \label{sec:data_methods}


In this study, we investigate the spatially resolved optical emission lines of X-ray selected AGNs that have unusual emission line properties in SDSS. In particular, we wish to evaluate to what extent the measurement of an AGN's optical emission lines may be affected by two types of aperture effects: due to continuum swamping from particularly bright hosts, and dilution of emission lines due to star formation.

To do so, we select AGNs using the method of \cite{agostino2023} which identifies AGNs as the galaxies which have X-ray emission in significant excess compared to what is expected based on their star formation rate (SFR). We then use the MPA/JHU catalog of quantities derived from 3\arcsec\ SDSS spectroscopy \citep{tremonti2004} to identify a sample of X-ray-selected AGNs which lack AGN-like emission lines.

In the following subsections, we describe the data sources from which we select our weak-line AGN sample (Section \ref{sec:dm_data}), the use of optical emission line diagnostics we employ to select our parent sample (Section \ref{sec:sample}), a description of the targets in the parent sample that we selected for follow-up observations with VLT-MUSE  (Section \ref{sec:musedata}), and a description of the method used for correcting optical emission lines for dust extinction (Section \ref{sec:dust}).

\subsection{Data Sources}
\label{sec:dm_data}

To carry out the selection of X-ray AGNs, we use sources from the tenth release of the fourth XMM-Newton serendipitous source catalog (4XMM, \citealt{webb2020}). We follow the same X-ray data processing steps as outlined in detail in \citet{agostino2023}. In summary, we derive hard-band (2-10 keV) and full-band (0.5-10 keV) X-ray luminosities from X-ray fluxes, assuming a power-law spectral model with a photon index $\Gamma=1.7$. We retain only those sources with S/N$>2$ in the hard band. We additionally utilize the extent of the X-ray sources---available in the `ext' column of the 4XMM catalog---to distinguish between resolved and unresolved X-ray sources, and retain only those sources which are unresolved. Unresolved sources are more likely to be genuine AGNs, whereas resolved sources likely owe their X-ray emission to hot gas. 

We require SFRs in order to use the X-ray excess method from \citet{agostino2019} and \citet{agostino2023} for identifying X-ray AGNs. We obtain SFRs, along with stellar masses and stellar continuum dust attenuations, from the medium-deep UV survey of the GALEX-SDSS-WISE Legacy Catalog (GSWLC-M2\footnote{https://salims.pages.iu.edu/gswlc/}; \citealt{salim2016, salim2018}). These parameters were determined via the UV/optical+IR SED fitting. Galaxies in GSWLC-M2 were matched to X-ray sources following the procedure in \citet{agostino2023} whereby we use a 7\arcsec search radius. If there are multiple GSWLC-M2 galaxies that match to a single X-ray source, we retain only the galaxy with the brightest SDSS $r$-band magnitude. The spatial resolution of the X-ray imaging is 6\arcsec.

We use optical emission line fluxes from the MPA/JHU catalog derived from 3\arcsec\ SDSS spectroscopy following \citet{tremonti2004}.

\subsection{Optical emission line selection} \label{sec:sample}


To select a parent sample of objects for which to test aperture effects like host dilution from star formation or continuum swamping by bright hosts, we select X-ray AGNs using the X-ray excess method. Following \citet{agostino2019}, we compare the measured X-ray luminosity of galaxies to the amount expected for the galaxy given its SFR and the empirical relationship between X-ray luminosity and SFR for non-AGNs, as measured by \citet{ranalli2003}. In so doing, we identify $\sim$500 X-ray AGNs which comprise our parent sample. From this parent sample, we use optical emission to select a subset of objects as potential targets. 

Practically, we select as potential targets X-ray AGNs with S/N$<2$ in one or more of the BPT emission lines (we refer to these as weak-line X-ray selected AGNs or WL-XAGNs) or with high S/N emission lines that place the X-ray AGN within the star-forming region of the BPT diagram (we refer to those as star-former X-ray AGNs or SF-XAGNs). 

Explicitly, we define the star-forming region of the BPT diagram as the portion lying below a modified version of the \citet{kauffmann2003} line---which is the original \citet{kauffmann2003} line for log([NII]/H$\alpha)<-0.4$, at which point it becomes a one-dimensional boundary. Galaxies that lie above the \citet{kauffmann2003} line for log([NII]/H$\alpha)<-0.35$ or which have log([NII]/H$\alpha)>-0.35$ are considered BPT AGNs. These modifications to the \citet{kauffmann2003} demarcation line provide a somewhat cleaner separation of non-AGNs and AGNs (as described in \citealt{agostino2021}) and the modified demarcation line (with the vertical AGN boundary at log[NII]/H$\alpha=-0.35$) ) is shown in our figures analyses wherever a BPT diagram appears.

Using this scheme, we identify $\sim$40 SF-XAGNs and $\sim$130 WL-XAGNs for our target sample.

\subsection{Targets of interest} \label{sec:target}
 To select targets of interest for follow-up observation, we first considered observing constraints from the southern hemisphere and then prioritized the observable SF-XAGNs and WL-XAGNs based on their distance from the Ranalli relation.
 We selected 9 targets (4 SF-XAGNs and 5 WL-XAGNs)  from our parent sample for follow-up observations.

In carrying out our parent and target sample selection, we used an earlier version of the XMM-Newton serendipitous source catalog (3XMM-DR6). In the analyses we perform in this paper, we use an updated version (4XMM-DR10) of the X-ray catalog and in this updated version, one of our WL-XAGN sources is no longer considered an unresolved X-ray source and so it is probably not likely to be a genuine X-ray AGN. To differentiate between this extended X-ray source and the others, we designate it as WL-EXT-1. 

\subsection{VLT-MUSE Observations} \label{sec:musedata}

We obtained observations of 5 of the 9 targets in the 2020B Semester as part of Program 106.21CH; the other 4 targets were not observed due to queue pressures. The observations were carried out in the Wide Field Mode (WFM) with adaptive optics on the Multi-Unit Spectroscopic Explorer (MUSE) integral field unit (IFU) spectrograph  on the Very Large Telescope (VLT), which we will hereafter refer to as VLT-MUSE. 

Each spaxel has a size of 0\farcs2 by 0\farcs2. The characteristic FWHM is 0\farcs{3} ---0\farcs{4} and so the improvement in spatial resolution over SDSS 3\arcsec\ fiber spectroscopy is $\sim$10 times.  The wavelength range in each data cube spans $\sim$4750---9350 \AA\ with a spectral sampling of 1.25 \AA\ and a $\sim$2.5 \AA\ FWHM  ($\sim$50 km s$^{-1}$) around the H$\alpha$ emission line. The spectral resolution ranges from 1750 (blue end) to 3750 (red end), a substantial increase over SDSS which ranges from 1500 (blue end) to 2500 (red end). 

Exposure times (Table \ref{tab:obs}) for WL-XAGNs were determined with the VLT-MUSE exposure time calculator so as to obtain S/N $\sim10$ in the blue part of the spectrum. For the SF-XAGNs, a lower S/N threshold (S/N$>5$) in the blue part of the spectrum was targeted. 

\begin{figure}
         \begin{center}
            \includegraphics[width=\columnwidth]{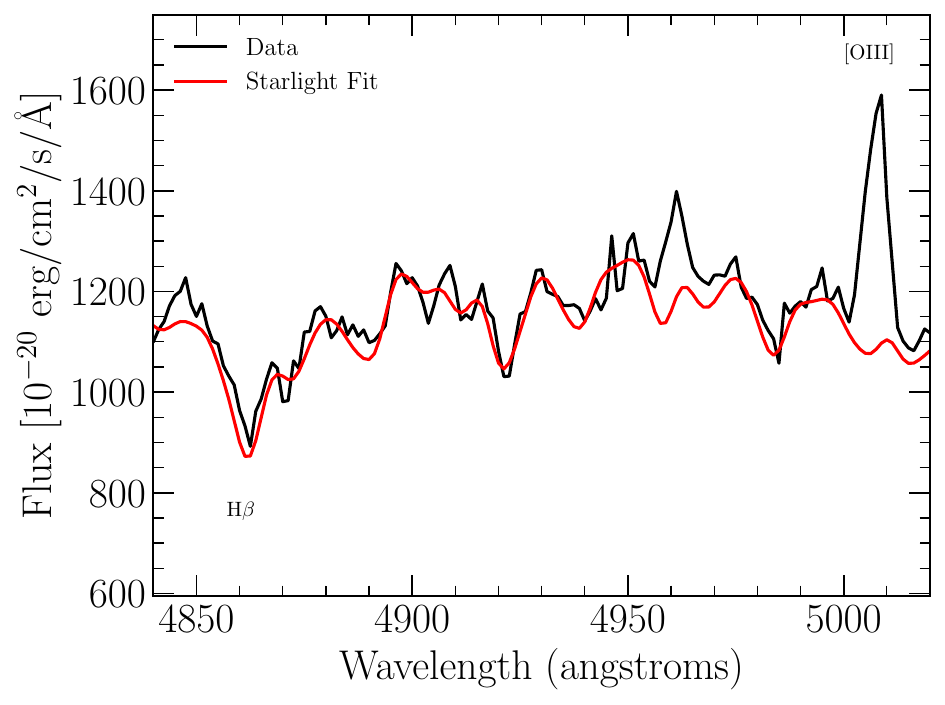}
            \includegraphics[width=\columnwidth]{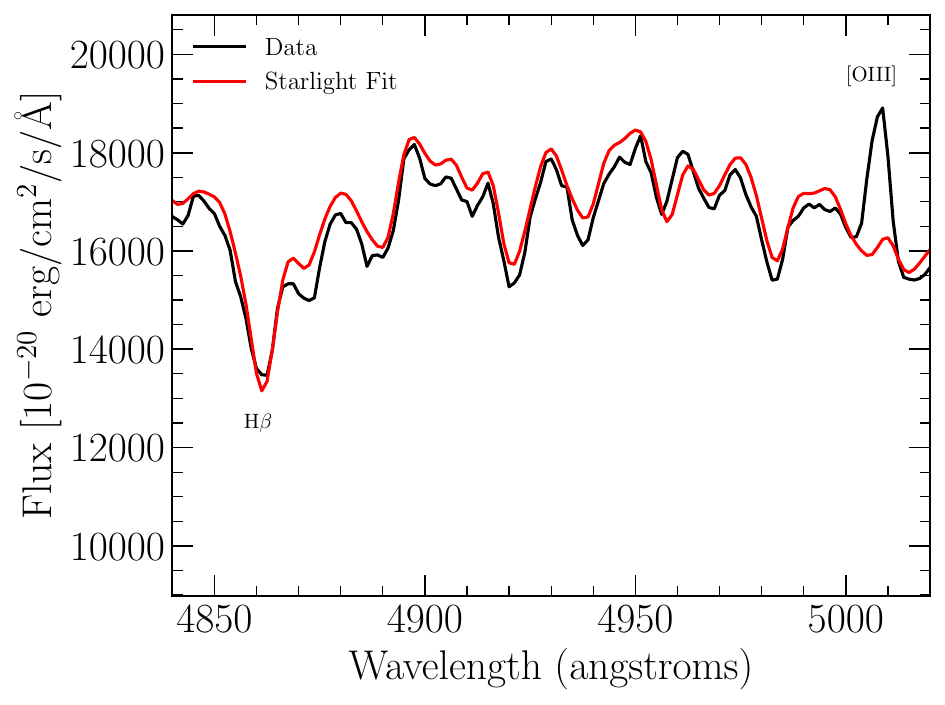}
            \caption{MUSE spectra of WL-1 in 0\farcs 5 (top) and 3\arcsec (bottom) apertures. MUSE spectra are shown in black and \texttt{STARLIGHT} fits are shown in red. Labels are also included to indicate the positions of the H$\beta$ and [OIII] lines.
            \label{fig:specs}}
        \end{center}
        \end{figure}

We reduced, sky subtracted, and extracted the VLT-MUSE data into data cubes using the MUSE data pipeline \citep{muse_pipeline}. 
 We used the \texttt{MPDAF}\footnote{https://mpdaf.readthedocs.io/en/latest/index.html} Python library to load and inspect our data and \texttt{STARLIGHT} to fit and subtract the stellar continuum in each spaxel. In the fitting process, we used 45 stellar populations from \citet{bc2003} with 15 ages (ranging from 3 Myr to 13 Gyr) and 3 metallicities ($Z/Z_{\odot}=$[0.004, 0.02, 0.05]) to model the continuum and found that using more (40 ages and 6 metallicities) would not alter the result while significantly increasing the computation time. In Figure \ref{fig:specs}, we provide examples of the spectral and continuum quality for WL-1 in the wavelength range surrounding H$\beta$ and [OIII], as these are typically the noisiest lines.

 \begin{table*}
 \begin{tabular}{c|ccccccc}
 \hline
 Object &  Name & $t_{\mathrm{ exp}}$ (total) &Observation Date(s) & $N_{\mathrm{exp}}$ & Physical resolution limit & redshift  & $A_{V, \textrm{gas}}$\\
 & & s & (YYYY-MM-DD) &  & kpc & z & \\
 \hline
SF-1 &SDSS J095513.89+173703.7 & 5529	 & 2020-12-09  2020-12-21  & 3 & 0.56 & 0.071 &1.2\\
WL-1 & SDSS J112044.00+125031.3 & 3064 & 2020-12-04, 2020-12-14, 2020-12-21  &4 & 0.76 & 0.101 & 0.86\\
WL-2 & SDSS J010332.35-001849.9 & 450 & 2021-01-03 & 3 & 0.72 & 0.096 & 0.29\\
WL-3& SDSS J010155.76+002816.3 & 840  & 2021-01-14& 3 & 0.72 & 0.095 & 0.54\\
WL-EXT-1 & SDSS J090256.02+014626.0 & 1158  & 2020-11-13 & 3 & 0.88 & 0.118 & 0.55\\
\hline
\end{tabular}
\caption{Observing log of MUSE observations and properties of galaxies. Physical resolution limits are determined by multiplying the angular scale (in kpc/\arcsec) by the characteristic FWHM of VLT-MUSE, here assumed to be 0\farcs{4}.}
\label{tab:obs}     
 \end{table*}

Continuum-subtracted datacubes were fit using the \texttt{DOBBY}\footnote{DOBBY is available at https://bitbucket.org/streeto/pycasso2.} suite of procedures \citep{asari2019} that are available as part of PyCasso \citep{cidfernandes2013, deamorim2017}. \texttt{DOBBY} uses a Legendre polynomial to model any remaining underlying (smoothly varying) continuum and simultaneously fits Gaussian profiles to a suite of optical emission lines, including but not limited to the BPT emission lines, [OI], and [SII]. Negative fluxes for emission lines were not permitted in our fitting procedure. Lines of the same species (e.g., H$\alpha$ and H$\beta$) were kinematically tied during the line fitting process and the Balmer decrement (H$\alpha$/H$\beta$) was further constrained to be above 2.6. Emission line flux errors were determined following \citet{rola1994} and \citet{asari2019}: 
\begin{equation}
    \sigma_{F} = \sigma_{N} \sqrt{6\sigma_{\lambda}\Delta\lambda}
\end{equation}
where $\sigma_{N}$ is the RMS in the detrended residual continua in windows blueward and redward of the lines, $\sigma_{\lambda}$ is the Gaussian dispersion in \AA,  and $\Delta \lambda$ is the spectral sampling (1.25 \AA). 

For directly measuring the aperture effects, we extract spectra in various aperture sizes ranging from 0\farcs5 (4 spaxels) to 3\arcsec\ (177 spaxels), in steps of 0\farcs5, using the aperture extraction tool in MPDAF.
Aperture spectra were summed, analogous to what would have been obtained with single fiber spectroscopy.
We carried out an exercise wherein aperture extraction was done before continuum subtraction. We found little difference between the two results. To verify that the above procedure was not being affected by any systematic issues, we also took the SDSS spectrum of each galaxy and used \texttt{STARLIGHT} to remove the stellar continuum and \texttt{DOBBY} to fit the emission lines of the residual spectrum. We compared the emission line ratios measured in the MPA/JHU catalog to those measured with \texttt{DOBBY} and found no systematic differences. The fluxes of the BPT lines measured by MPA/JHU and \texttt{DOBBY} on the SDSS spectrum are given in Table \ref{tab:fluxes}. together with the MUSE fluxes in 0\farcs5 and 3\arcsec\ apertures.

\begin{table*}
  \begin{tabular}{c|cc|cc|cc|cc|cc|c}
  \hline
      Line& Source & \multicolumn{2}{c}{SF-1} &\multicolumn{2}{c}{WL-1} &\multicolumn{2}{c}{WL-2} &\multicolumn{2}{c}{WL-3} &\multicolumn{2}{c}{WL-EXT-1} \\
      & &Flux &SNR &Flux &SNR &
     Flux &SNR &Flux &SNR &
     Flux &SNR        \\
\hline
[OIII] & a &
    20 & 9 & 
    13 & 7.4 & 
    24 & 6.3 & 
    0 & 0 &
    0 & 0   \\
    & b  &
    21 & 8.3 & 
    15 & 6.6 &
    37 & 6.8 &
    2 & 0.9 &
    $-2$ & $-1.2$      \\
 & c&
 18 & 43 & 
 18 & 39 &
 36 & 14.7 &
 2.1 & 3.4 & 
 0 &0 
    \\
 & d&
 0.7 & 12 &
 2.8 & 23 &
 4 & 7.3 & 
 0.2 & 0.8 &
 0 & 0     \\
\hline
 [NII] & a&
    99 & 58 &
    24 & 11.6 &
    67 & 13 & 
    6 & 0.61 & 
    0 & 0  
        \\
 & b&
 98 & 37 &
 31 & 12 &
 109 & 18 &
 1 & 0.7 & 
 3 & 1.4  
     \\
 & c&
 107 & 180 &
 39 & 30 &
 133 & 31 & 
 0 & 0 & 
 10 & 4.2  
   \\
 & d& 
 3.7 & 117 &
 6.2 & 38 &
 17 & 30 &
 0 & 0 & 
 0.1 & 0.8 
  \\ 
\hline
H$\alpha$ & a &
282 & 165 &
18 & 9 &
38 & 7.3 &
6 & 0.6 & 
0 & 0  
   \\
 & b&
 287 & 71 &
 22 & 10 &
 66 & 9.8 & 
 $-10$ & $-1.8$ &
 4 & 1.8  
     \\
 & c &
 310 & 529 &
 28 & 21 &
 82 & 19 &
 0 & 0 &
 24 & 1.9   
     \\
 & d &
 12 & 369 &
 3.3 &22 &
 8 & 13 &
 0 & 0 &
 0.4 & 2.7  
   \\
\hline
H$\beta$ & a &
65 & 25 &
3 & 1.2 &
6.2 & 1.7 &
0.3 & 0.1 &
0 & 0 
   \\
 & b &
 63 & 23 &
 4 & 2 &
 3 & 0.5 &
 5 & 0.9 &
 $-5$ & $-2$.6  
    \\
 & c &
 69 & 163 & 
 5 & 6.4  &
 10 & 4.7 &
 0 & 0 & 
 1 & 1.6  
   \\
 & d &
 2.7 & 51 &
 0.2 & 2.2 &
 0 & 0 &
 0 & 0 &
 0.07 & 0.5  
    \\ 
 \hline
  \end{tabular}
      \caption{Table of observed fluxes and SNRs for       each of the 4 BPT lines in a) SDSS(our analysis), b) SDSS (MPA/JHU), c) MUSE 3\arcsec, d) MUSE 0\farcs5. Fluxes are in units of 10$^{-17}$  erg s$^{-1}$  cm$^{-2}$.    \label{tab:fluxes}}
 \end{table*}

\subsection{Dust Corrections}  \label{sec:dust}

Deriving reliable dust corrections with the Balmer decrement method requires both H$\alpha$ and H$\beta$ to be well-measured and accurate. Thus the continuum surrounding the H$\beta$ line must be fit well. Otherwise, given that a noisier continuum will tend to wash out potential emission from H$\beta$, H$\beta$ will be underestimated. 

When the S/N in SDSS H$\beta$ is larger than 10, we correct the optical emission for the dust extinction using the Balmer decrement method. We assume the dust-free ratio of 3.1 for narrow-line regions \citep{agn_squared}  and the \citet{cardelli1989} attenuation curve. 
 When the S/N of H$\beta$ is $\leq 10$, the Balmer decrement is poorly determined and so, following the process described in \citet{agostino2021}, we estimate gas-phase attenuation from stellar continuum dust attenuation ($A_V$ from GSWLC-M2 based on SED fitting). The estimation is based on the relationship between the Balmer decrement and $A_V$ for objects with well-determined Balmer decrement (S/N in H$\beta>10$). We apply the same correction for a given target for all observations (SDSS and MUSE), and include the gas phase attenuation in Table \ref{tab:obs}


\begin{figure}
         \begin{center}
            \includegraphics[width=\columnwidth]{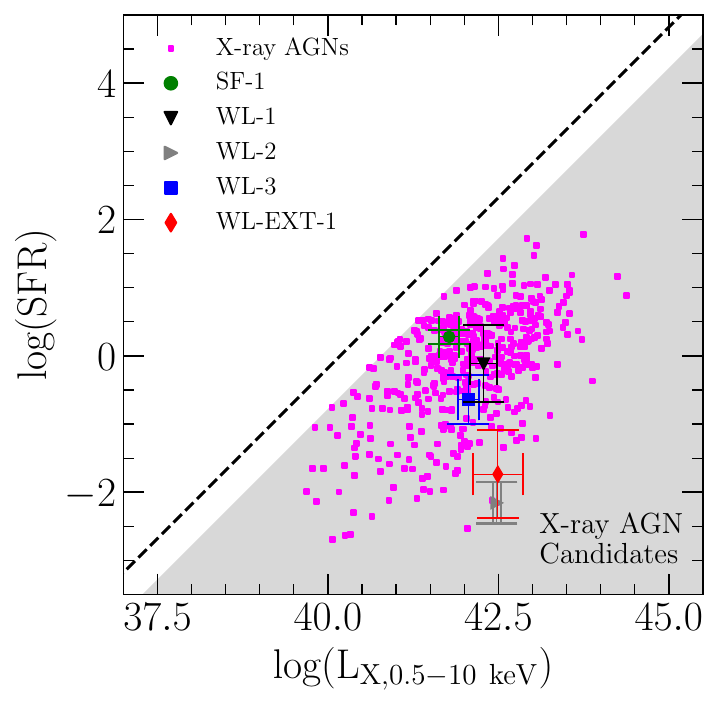}
            \caption{Star formation rate versus X-ray luminosity for VLT-MUSE targets (large symbols) and the parent sample of X-ray AGNs (small magenta squares). Error bars are shown for the MUSE targets. Shaded region indicates how X-ray AGN candidates are selected based on X-ray excess with respect to SFR. Dashed line is the L$_{X}$-SFR relation for non-AGN galaxies originally from \citet{ranalli2003} and adapted by \citet{agostino2019}. \label{fig:lxsfr}}
        \end{center}
        \end{figure}

\begin{figure}
         \begin{center}
            \includegraphics[width=\columnwidth]{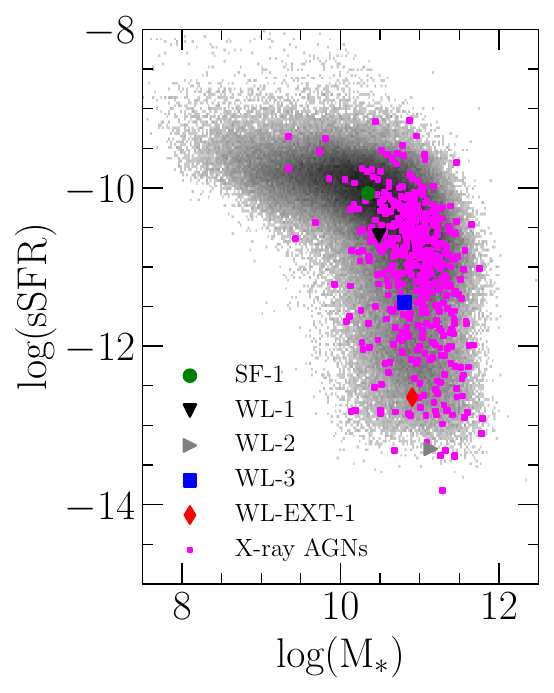}
            \caption{Specific star formation rate versus stellar mass for VLT-MUSE targets and the parent sample of X-ray AGN candidates. We show a two-dimensional histogram of the SDSS distribution of galaxies and the shading is determined by the number density to the 1/3 power so as to highlight outliers. \label{fig:ssfrm}}
        \end{center}
        \end{figure}

\section{Results} \label{sec:results}

We first provide an overview of the different types of analysis being performed, and then discuss each of the targets in the context of this analysis.

We compare the full-band X-ray luminosity and SFR (Figure \ref{fig:lxsfr}) and the sSFR and stellar mass (Figure \ref{fig:ssfrm}) of the VLT-MUSE targets and the parent sample. In these comparisons, we also make use of the general population of galaxies that are available in GSWLC-M2.

We present maps derived from MUSE observations for five target galaxies in Figures \ref{fig:SF-1}-\ref{fig:WL-1}: we show an RGB image of each galaxy based on broadband fluxes extracted from a spectrum before the removal of its continuum, a BPT classification map (where each spaxel is assigned either as star-forming or as AGN based on the position on the BPT diagram), the fluxes in each of the four BPT emission lines, and the line ratios of [OIII]/H$\beta$ and [NII]/H$\alpha$. In each of the flux maps, a spaxel is shown only if a line has S/N$>2$. In the line ratio plots, the spaxels are shown if both emission lines have S/N$>2$. In the BPT classification map, all four lines must have S/N$>2$ in order for a spaxel to be shown. 

We show BPT diagrams based on the different apertures we extract (Figure \ref{fig:bpts}). 
The line ratios of apertures are shown if they have all 4 BPT lines with S/N$>2$. For the line ratios derived from aperture-summed spectra, we color-code the data based on the S/N of the BPT emission lines: the circles are colored orange if they have S/N$>2$ in all four BPT line and are colored red when either [OIII] or H$\beta$ have S/N$<2$ but [NII] and H$\alpha$ are well detected (both with S/N$>2$). In the latter case, we plot the [OIII]/H$\beta$ value based on the smallest aperture where both lines are detected. In these plots, we show the overall distribution of SDSS optical emission line ratios for GSWLC-M2 galaxies.

We consider the properties of the target sample in the context of the relationship between the X-ray and [OIII] emission of AGNs (e.g., \citealt{heckman2005, panessa2006}) in Figure \ref{fig:lxo}. 

In what follows, we discuss each object individually but focus on the two groups (SF-XAGNs and WL-XAGNs) and the particular scenarios that have been proposed to explain their properties.


 \subsection{SF-XAGNs: AGNs with HII-like lines} \label{sec:xrsf}
 
 For the X-ray AGNs which have been classified as star-forming based on their optical emission lines, we have three primary scenarios to explain their apparent misclassification.
 \begin{enumerate}
     \item AGN lines of normal strength have been diluted (actually overwhelmed) by host star formation such that their resulting emission line classification places it in the SF region of the BPT diagram instead of the AGN branch \citep{moran2002}.

     \item AGN lines are weak, so we only see lines from host SF \citep{agostino2019}.
     
     \item The AGN has a soft enough ionizing spectrum that it effectively presents itself with HII-region-like emission. 
     \\
 
 \end{enumerate}

IFU spectroscopy allows us to mitigate host contamination, allowing us to distinguish between scenarios (i) and (ii). We can also use IFU spectroscopy to investigate the likelihood of scenario (iii) by studying the spatial distribution of the emission line ratios.

         \begin{figure*}
         \begin{center}
            \includegraphics[width=2\columnwidth]{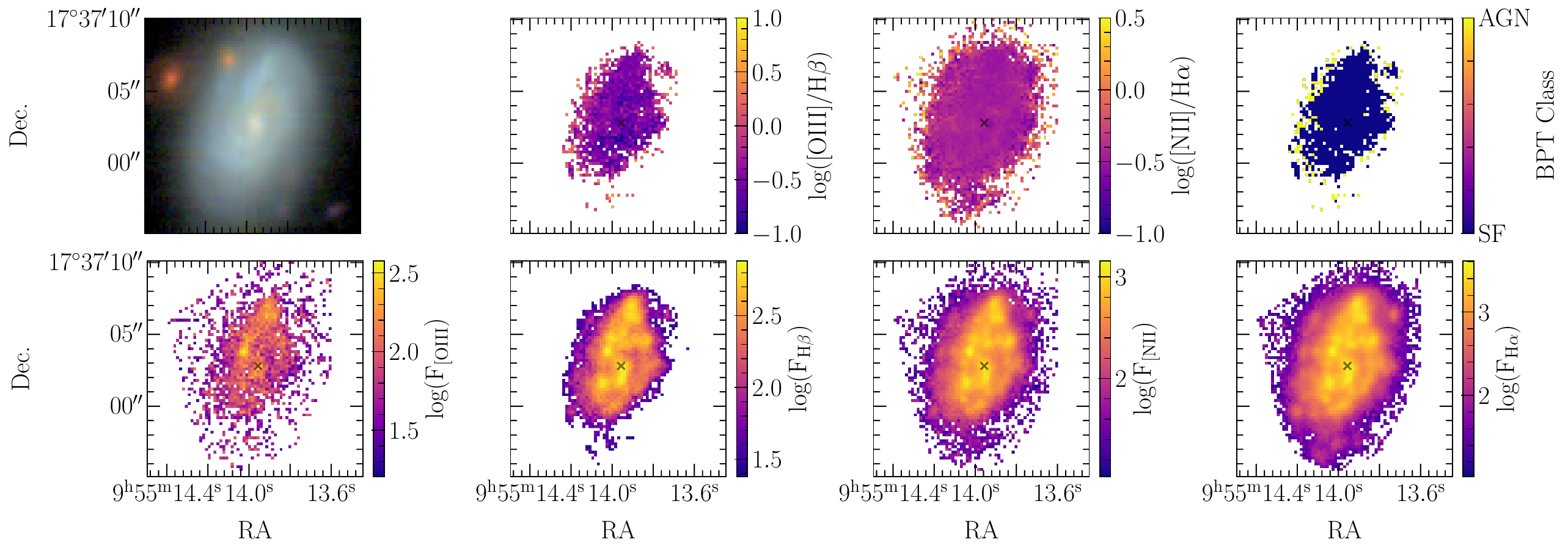}
            \caption{MUSE maps of SF-1. In the first row from left to right: RGB image constructed from MUSE spectroscopic data, log([OIII]/H$\beta$), log([NII]/H$\alpha$), and BPT class. In the second row from left to right: the flux in [OIII], the flux in H$\beta$, the flux in  [NII], and the flux in H$\alpha$. \label{fig:SF-1}}

        \end{center}
        \end{figure*}

 Visually, SF-1 is a spiral galaxy (Figure \ref{fig:SF-1}) with extensive ongoing star formation in its disk. Compared to other X-ray AGNs, SF1 has a low L$_{X}$/SFR ratio and mass and a high sSFR (Figures \ref{fig:lxsfr} and \ref{fig:ssfrm}). 
 
 In the maps of its ionized gas emission (Figure \ref{fig:SF-1}), there are clear clumpy structures associated with spiral arms. Most of SF-1 has a roughly constant [NII]/H$\alpha$---a metallicity indicator for star-forming galaxies \citep{vanzee1997}. [OIII]/H$\beta$ shows more variation but it does not increase towards the central parts of the galaxy as one might expect from optical emission lines that are produced as a result of AGN activity. There do not appear to be any AGN emission lines in the central region of the galaxy even at the high spatial resolution of 0\farcs{3}\ or $\sim$ 0.4 kpc at z=0.0712). 
 For context, the size of the narrow-line region of an AGN will depend on how luminous the AGN is \citep{bennert2006a_nlr, bennert2006b_nlr, greene2011_nlr, liu2013_nlr, liu2014_nlr, law2018_nlr, chen2019nlr}, and may range from 1-10 kpc. 
 
  Its measured line ratios in the various apertures (Figure \ref{fig:bpts}) are all concentrated around the same region and so the inclusion of extra light in the aperture does not appear to be changing the result. Interestingly, the star formation rate for SF-1 determined from the H$\alpha$ emission is consistent ($<0.1$ dex difference) with that from the SED, further suggesting there is no hidden AGN component altering the measurement.

 One might argue that SF-1's line ratios could be the result of ionization by a weaker AGN (scenario iii) that effectively mimics the line ratios of HII regions. However, given that there is large homogeneity in the BPT line ratios across the galaxy and far outside of where the NLR might be expected, it seems more likely that the dominant ionization source throughout is the star formation.

 Based on the \citet{panessa2006} relation between X-ray and optical emission, the expected [OIII] luminosity of SF-1 should be log(L$_{\mathrm{[OIII]}}$)$\approx$40.4 dex and based on the relation between [OIII] luminosity and NLR size from \citet{chen2019nlr} (defined as R$_{16}$ which is the size of the NLR at a surface brightness cut of 10$^{-16}$ erg s$^{-1}$ cm$^{-2}$ arcsec$^{-2}$), the NLR of SF-1 should have a diameter of $\sim$2 kpc and so should be detectable across multiple spaxels if it were indeed dominant. 
 The fact that an AGN signature is still not found suggests that AGN lines must be intrinsically weak, ruling out the SF dilution explanation (scenario i).
Its lack of AGN-like signature in its 0\farcs5 spectrum shows that whatever its intrinsic [OIII] is, it must be quite underluminous.

 
  In summary, SF-1's AGN lines are intrinsically weak and are not simply overwhelmed by SF. As a matter of fact, if there was no SF, this object would belong to our second category of objects that we explore in this paper and thus scenario (ii) is the most likely of the explanations. 
 
 \subsection{WL-XAGNs: AGNs with weak or no emission lines}
 
 For the X-ray AGNs which have weak or no lines in the SDSS spectroscopy, there are a variety of potential scenarios:
\begin{enumerate} 
\item lines are made weak because they are swamped by the continuum light from the host. While also proposed by \citet{moran2002}, this scenario is distinct from scenario (i) for AGN with SF lines (Sec \ref{sec:xrsf}), where host contamination was in line emission.
\item lines are obscured by nuclear dust \citep{moran1996, comastri2002} or host dust \citep{rigby2006}.
\item lines are intrinsically weak because of the lower production of UV photons \citep{yuan2004}.
\item the absorption of ionizing photons by the NLR is inefficient either because of a complex geometry \citep{trouille2010} or because the ISM is gas-poor \citep{herpich2018, agostino2023}
\end{enumerate}  
 IFU observations are again well poised to distinguish between the popular scenario (i) (host swamping) and others.

         \begin{figure*}
         \begin{center}
            \includegraphics[width=2\columnwidth]{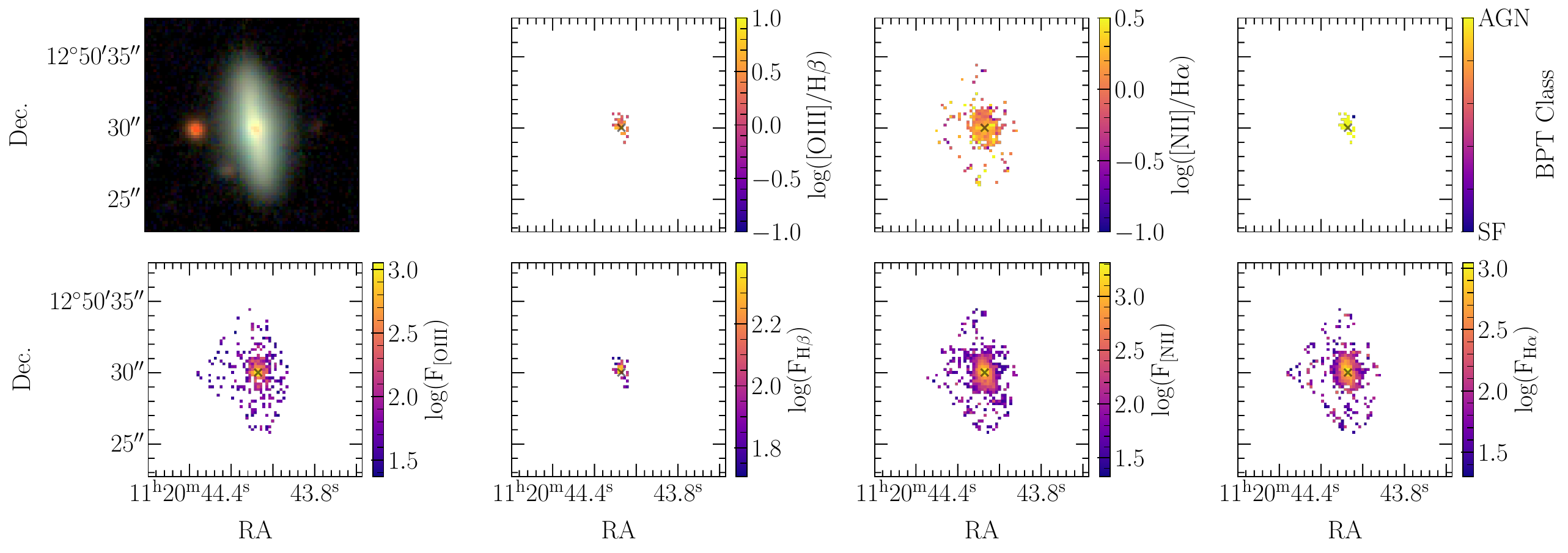}
            \caption{MUSE maps of WL-1.  In the first row from left to right: RGB image constructed from MUSE spectroscopic data, log([OIII]/H$\beta$), log([NII]/H$\alpha$), and BPT class. In the second row from left to right: the flux in [OIII], the flux in H$\beta$, the flux in  [NII], and the flux in H$\alpha$.\label{fig:WL-1}}

        \end{center}
        \end{figure*}

 
 \subsubsection{WL-1}
 Visually, WL-1 is an edge-on S0. WL-1 has an L$_{X}$/SFR ratio typical of other X-ray AGNs (Figure \ref{fig:lxsfr}). Its mass is on the low side, and its mass and sSFR place it on the lower side of the main sequence (Figure \ref{fig:ssfrm}). 
 
 The VLT-MUSE maps for WL-1 are shown in Figure \ref{fig:WL-1}. One can clearly see strong optical emission in its central region. [OIII], [NII], and H$\alpha$ are well measured among many spaxels whereas H$\beta$ is poorly measured outside of the central region. For the spaxels which can be classified, they are well within the AGN region of the BPT diagram and cluster around the same [NII]/H$\alpha$ ratio as measured from SDSS (Figure \ref{fig:bpts}). The 3\arcsec\  SDSS and MUSE fluxes are consistent but the SNRs for the MUSE fluxes are higher because of the increased depth. 
  
 The strength of WL-1's emission decreases radially as would be expected from an ionization process like a central AGN. As the aperture size increases, we can see this effect directly: the [OIII]/H$\beta$ value decreases by $\sim$0.4 dex and the [NII]/H$\alpha$ by 0.1 dex (Figure \ref{fig:bpts}), a shift that is not atypical for aperture effects \citet{agostino2021}. The initial [OIII]/H$\beta$ position of the 0\farcs 5 aperture is also subject to some uncertainty, having an error of 0.2 dex, primarily because of the low S/N in H$\beta$ in this aperture (S/N=2.2), so the weakening may not be as extreme as it appears. In any case, this `weakening' of the AGN relative to its smallest aperture may be due to the inclusion of more light from star formation. The galaxy should have some SF according to SED fitting estimates (Fig.~\ref{fig:ssfrm}). This result implies that many AGNs in SDSS might exhibit lower BPT line ratios than they possess nearer the central engine. This effect is likely most pronounced for the weaker AGNs as the more luminous ones will have a larger \textit{fully} ionized NLR. 
 
 As for its [OIII] luminosity, we find that the measured luminosity within the  0\farcs5 aperture is substantially lower than that measured within the 3\arcsec\ aperture. If dilution were occurring in the 3\arcsec\ aperture, we might expect the line to be easier to detect in the smaller aperture, but this is not obviously the case. Instead, the emission lines are intrinsically weak. Given its sSFR that is only somewhat lower than the main sequence ridge, it does not seem that the principal cause of it having a lower [OIII] luminosity is due to a lack of gas content, but it is possible that the covering factor of the NLR is below average and that the NLR emissions are diminished, as suggested by \cite{trouille2010}. Additionally, there may be an increased amount of dust surrounding the nucleus and so the NLR is perhaps not as efficiently heated as it could be if unimpeded.

        In summary, WL-1 does exhibit aperture effects in the sense that the position on the BPT diagram shifts towards the top of the AGN branch as the aperture is decreased. However, the principal reason for it being classified as weak-lined (low SNR) in SDSS is due to the AGN lines being intrinsically weak, rather than being overpowered by host light. Indeed, the SNR of [OIII] does not increase when the aperture is reduced.

     \begin{figure*}
        \begin{center}
            \includegraphics[width=2\columnwidth]{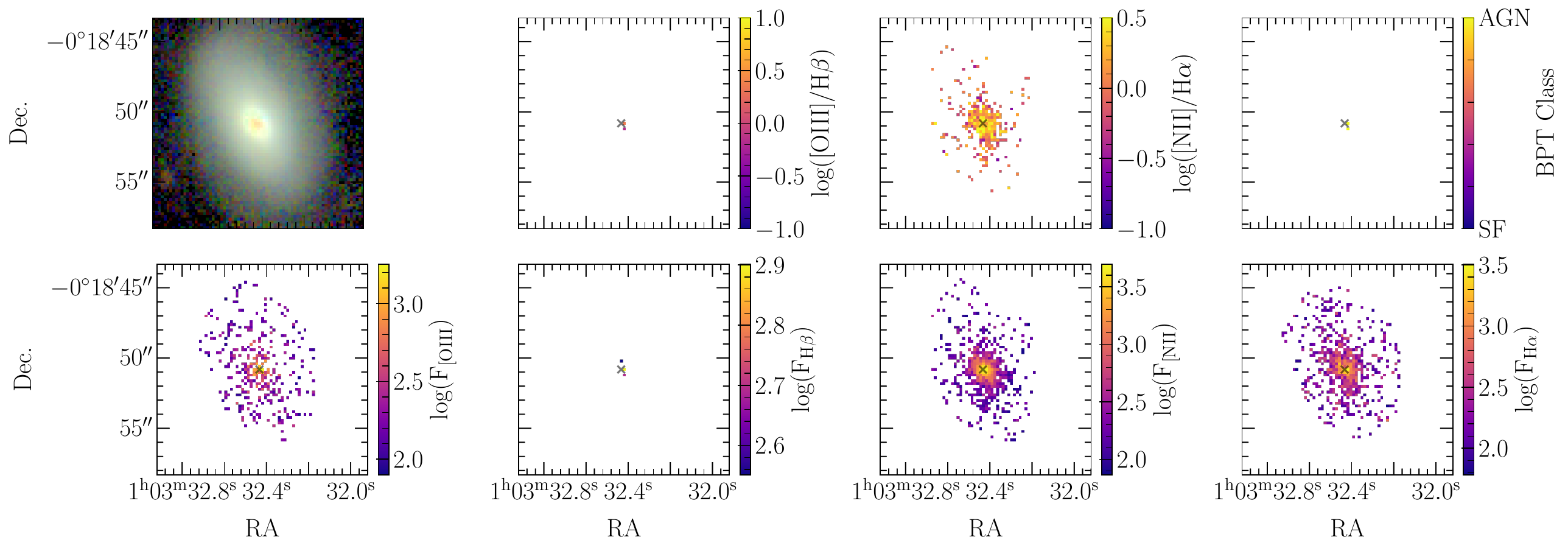}
            \caption{MUSE maps of WL-2. In the first row from left to right: RGB image constructed from MUSE spectroscopic data, log([OIII]/H$\beta$), log([NII]/H$\alpha$), and BPT class. In the second row from left to right: the flux in [OIII], the flux in H$\beta$, the flux in  [NII], and the flux in H$\alpha$\label{fig:WL-2}}

        \end{center}
        \end{figure*}
        
 \subsubsection{WL-2}

WL-2 is an early type galaxy. It has an extremely high L$_{\mathrm{X}}$/SFR ratio and an extremely low sSFR compared to the average X-ray AGN (Figure \ref{fig:lxsfr} and Figure \ref{fig:ssfrm}) or even compared to a general population of galaxies. Its mass is near the average for AGN host galaxies.
 
  The VLT-MUSE maps for WL-2 are shown in Figure \ref{fig:WL-2}. With the MUSE data, one can see that WL-2 possesses strong optical emission in its central regions. [OIII], [NII], and H$\alpha$ are well measured among many spaxels but H$\beta$ remains the limiting factor for BPT classification outside of the very central region. In fact, WL-2 could have been classified as an AGN based on log([NII]/H$\alpha$) alone, being greater than the cutoff of $-0.35$.
  
  In the maps for WL-2, the strength of the emission from ionized gas in their central regions decreases radially as one would expect when the ionization is done by a central process like an AGN. Across the range of aperture sizes shown in Figure \ref{fig:bpts}, the [NII]/H$\alpha$ of WL-2 does not move substantially, $\sim0.1$ dex. The small shift is expected because there we expect no contribution from any HII regions. It is difficult to assess the effect on [OIII]/H$\beta$ because it is not as well measured at lower aperture sizes but from 2\arcsec\ to 3\arcsec\ the [OIII]/H$\beta$ is nearly constant.
  Based on the \lxo\ relation (Figure \ref{fig:lxo}),  we see that going to smaller apertures only makes the [OIII] become even more underluminous. This excludes the host dilution scenario and again points to intrinsically weak [OIII].

  In summary, WL-2 has a very low dust content and consequently a very low sSFR. Although its [NII] is relatively strong, its [OIII] is again quite underluminous for reasons unrelated to aperture effects Instead, it is consistent with the suggestion in \citet{agostino2023} that the amount of gas content plays a role in determining the luminosity.

        \begin{figure*}
        \begin{center}
            \includegraphics[width=2\columnwidth]{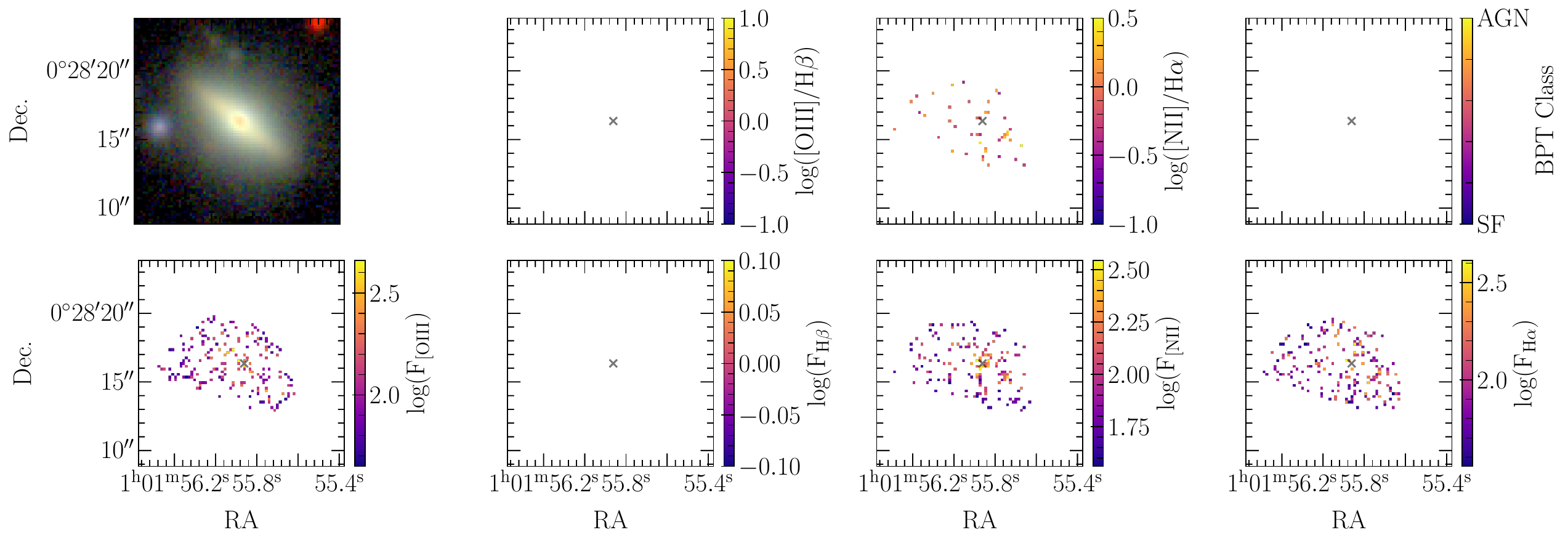}
            \caption{MUSE maps of WL-3. In the first row from left to right: RGB image constructed from MUSE spectroscopic data, log([OIII]/H$\beta$), log([NII]/H$\alpha$), and BPT class. In the second row from left to right: the flux in [OIII], the flux in H$\beta$, the flux in  [NII], and the flux in H$\alpha$\label{fig:WL-3}}
        \end{center}
        \end{figure*}
 
 \subsubsection{WL-3}
 WL3 is an edge-on S0 and has a L$_{X}$/SFR typical for X-ray AGNs  (Figure \ref{fig:lxsfr}). It has sSFR typical of a transitional galaxies, which have some gas and SF, but far below what is typical on the main sequence (Figure \ref{fig:lxsfr}).
 
 In the SDSS spectrum of WL-3, the BPT emission lines all have measured S/N levels less than 1. It is essentially a lineless galaxy, and the maps in Figure \ref{fig:WL-3} demonstrate that there is not a concentrated structure of ionized gas in the center of the galaxy, except for a nuclear patch of [NII] emission. 
 
 Emission lines remain un-detected in WL-3 in the smallest aperture. Thus, it is unlikely that the effect of continuum swamping is preventing the detection of the emission lines and instead that they are intrinsically quite weak. WL-3 has the weakest [OIII] luminosity in 3\arcsec\ aperture for its X-ray luminosity (Figure \ref{fig:lxo}). With even deeper spectroscopy, it is likely that an ionization structure like those seen in WL-1 and WL-2 would be found.

Another possibility is that the X-ray source is associated with a blue compact QSO candidate at z$\sim$0.8 that is 5\farcs 8 away \citep{yang2017}.  If this is the case, then the lack of emission lines in WL-3 will be not so surprising. 
 
 To summarize, WL-3 is either not a proper X-ray AGN, or is another example of an AGN that is powerful in the X-rays, but no matter how closely we isolate the nuclear region we do not find any emission lines, thus excluding the host swamping scenario. The host is an early type with little gas and not any obvious dust signature, which again suggests the picture in which the ionizing radiation simply does not have anything to ionize and we are left with a lineless AGN. Although WL-3 has weak lines, it is plausible that it belongs to the [OIII]-underluminous tail of the [OIII] luminosity distribution at its X-ray luminosity, like other XBONGs. 

         \begin{figure*}
         \begin{center}
            \includegraphics[width=2\columnwidth]{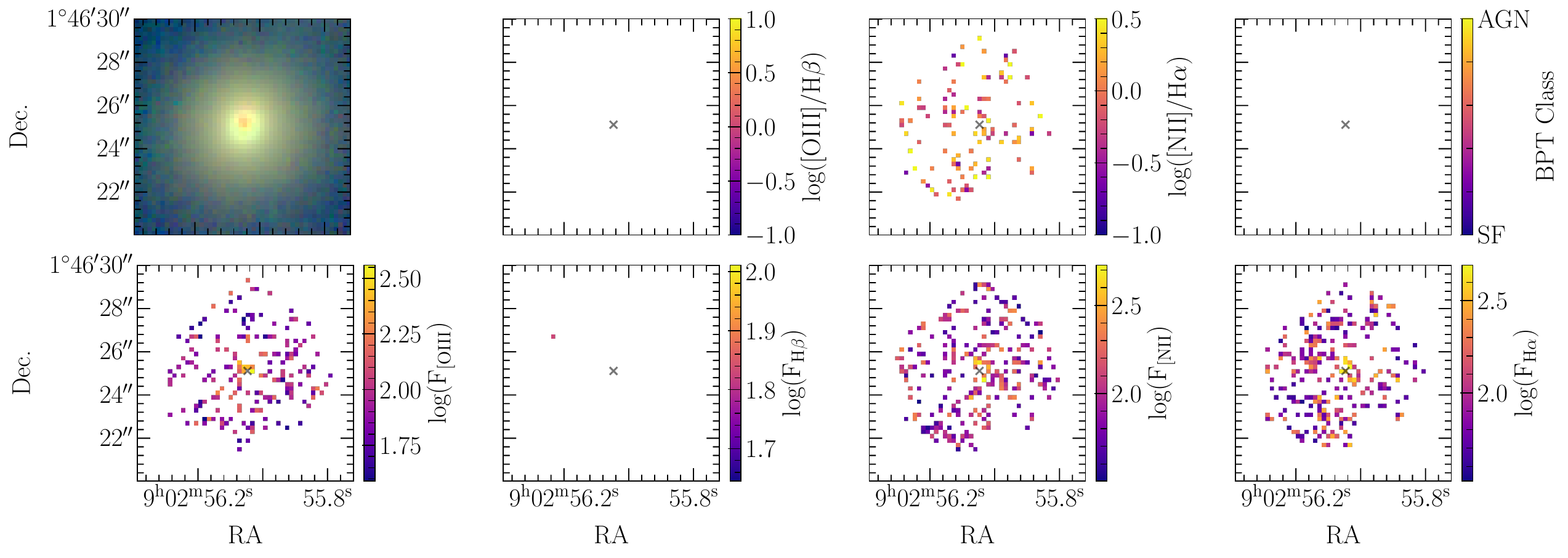}
           \caption{MUSE maps of WL-EXT-1. In the first row from left to right: RGB image constructed from MUSE spectroscopic data, log([OIII]/H$\beta$), log([NII]/H$\alpha$), and BPT class. In the second row from left to right: the flux in [OIII], the flux in H$\beta$, the flux in  [NII], and the flux in H$\alpha$\label{fig:WL-4}}

        \end{center}
        \end{figure*}
        
 \subsubsection{WL-EXT-1}
  Visually, WL-EXT-1 looks like an elliptical galaxy. It occupies a region where there are few X-ray AGN in L$_{X}$---SFR diagram (Fig 1). Its sSFR is consistent with being rather quiescent.
  
   It has some emission detected but none of the lines are detected in coincident spaxels so a full classification with the BPT diagram in any spaxel is impossible. In a visual assessment of the aperture spectra, they do not appear to show any genuine emission lines.

  WL-EXT-1 was initially included in this observational followup because 3XMM-DR6 had it as a point source \citep{agostino2023}, but in the more recent version (4XMM-DR10) it is listed as having a non-zero X-ray extent. While it is likely not associated with AGN activity, it provides an opportunity to investigate the optical emission line properties associated with extended X-ray sources. The VLT-MUSE maps for WL-EXT-1 are shown in Figure \ref{fig:WL-4}. It is worth to note here as well that this object is probably a member of a cluster of galaxies at $z=0.11$ \citep{bahar2022} and so the emission is likely associated with the hot gas in the intracluster medium. 

  %
  

To summarize, WL-EXT-1 is an elliptical galaxy with X-ray emission probably related to hot gas, not AGN. Spectroscopy supports that conclusion with no evidence of an AGN.

\begin{figure}
         \begin{center}
            \includegraphics[width=\columnwidth]{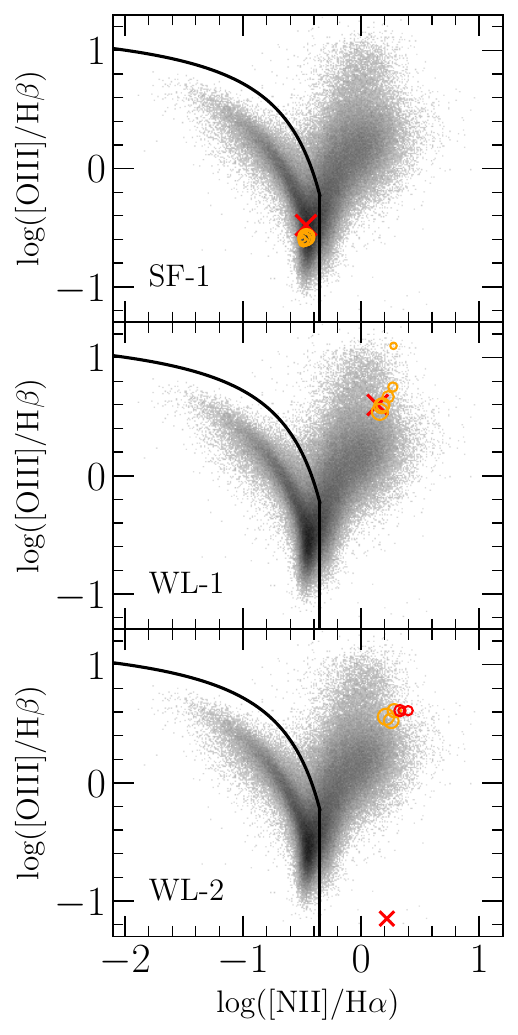}
            \caption{BPT diagrams for 3 X-ray AGNs with atypical optical spectra. We present the BPT line ratios as orange circles with their size a function of aperture size, increasing from 0\farcs5 to 3.0\arcsec.  When [OIII] or H$\beta$ has S/N$<2$, we colour the circles red and plot them at the [OIII]/H$\beta$ ratio of the smallest aperture which has S/N$>2$ in all four BPT lines.
            We show a two-dimensional histogram of the SDSS distribution of galaxies on the BPT diagram which have S/N$>2$ in all four lines (the shading is determined by the number density to the 1/3 power so as to highlight outliers), and plot the line ratios of the galaxy (red X) as available from the MPA/JHU catalog. In the case of WL-XAGN2, the H$\beta$ S/N is $<1$, but [NII]/H$\alpha$ is well-measured, and so we plot it at the measured [NII]/H$\alpha$ but place it arbitrarily at log([OIII]/H$\beta$) = $-1.2$
            The solid black line is the modified \citet{kauffmann2003} line which becomes a one-dimensional boundary at log([NII]/H$\alpha$)=$-0.35$ above which the objects are considered AGNs.
            WL-XAGN3 and WL-XAGN4 lack sufficient S/N ($>2$) in their aperture-derived BPT line ratios and so are not included in this plot.
            \label{fig:bpts}}
        \end{center}
        \end{figure}

\begin{figure}

         \begin{center}
            \includegraphics[width=\columnwidth]{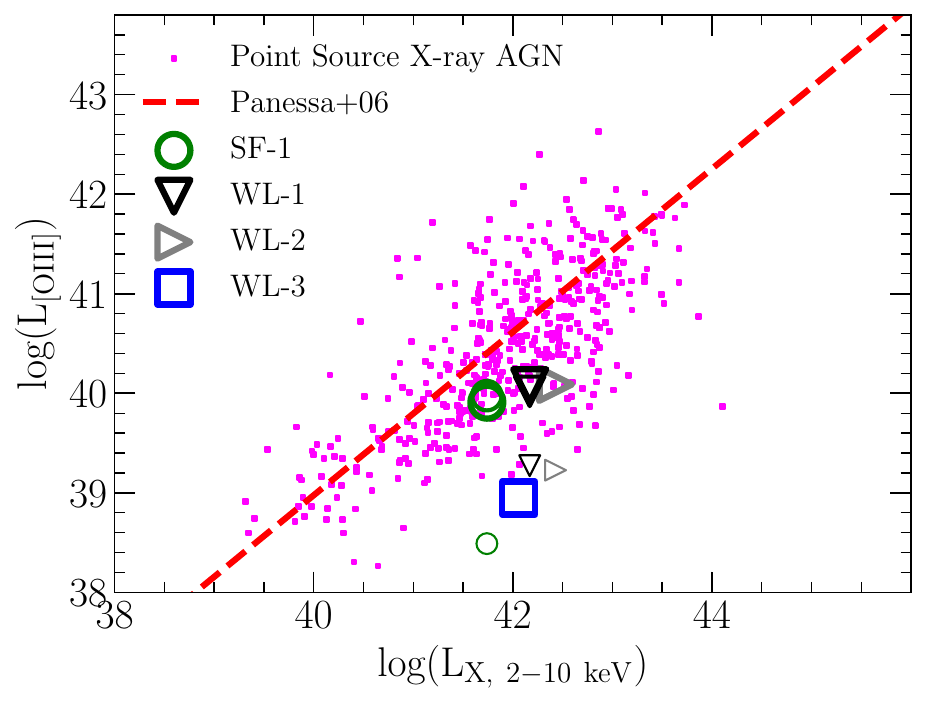}
            \caption{[OIII] luminosity versus X-ray luminosity for VLT-MUSE targets and comparison sample of X-ray AGN candidates. Dashed line is the relationship originally from \citet{panessa2006}. Magenta squares are X-ray AGNs. SF-1 is a green circle. 
            WL-1 is a black downward-facing triangle.
            WL-2 is a grey rightward-facing triangle.
            WL-3 is a blue square. For each galaxy, we show three measurements: 1) the MUSE 3\arcsec\ aperture-derived L$_{\mathrm{[OIII]}}$ (thickest border), 2) the MUSE 0\farcs5 aperture-derived L$_{\mathrm{[OIII]}}$ (thinnest border), and 3) the MPA/JHU 3\arcsec\ aperture-derived L$_{\mathrm{[OIII]}}$ (border thickness in between). 
            WL-3 does not have a valid measurement in its 0\farcs5 aperture or in the MPA/JHU catalog. 
            WL-EXT-1 is most likely not an X-ray AGN and has no measurable [OIII] and so is not shown in this plot.
            \label{fig:lxo}}
        \end{center}
        \end{figure}

\subsubsection{WL-XAGNs summary}
In summary, the three WL-XAGNs WL-1, WL-2, and WL-3 have AGN signatures and are consistent with being weak AGNs. They are not swamped by the continuum light of the host, disqualifying scenario (i). There is relatively little dust in any of the three AGNs so it does not seem plausible that scenario (ii) is viable either. This leaves us with scenarios (iii) and (iv) which are both still viable. The AGNs may be the result of a lower production of UV photons due to say a lower Eddington ratio but it is difficult to tell for certain given the uncertainties on measuring the black hole mass and in converting from X-ray luminosity to bolometric luminosity. On the other hand, it could be the case that ionizing photons are absorbed less efficiently either because of a complex NLR geometry or because the NLR does not have much gas, and our results are consistent with this picture. Both scenarios (iii) and (iv) are likely playing a part to some extent and disentangling them is critical for understanding the variety of observed AGN signatures.
    \section{Discussion} \label{sec:discussion}
In this study, we directly measured the impact of aperture effects on the measurements of AGN emission line ratios. As a result, we comment on the principal cause of the weak lines (Section \ref{sec:nature}) and the practical implications on optical emission line selection (Section \ref{sec:incompl}).

\subsection{Nature of weak line AGNs} \label{sec:nature}
In this paper, we did not find convincing evidence that aperture effects like continuum dilution or star formation dilution are able to explain the weak emission lines of some X-ray AGNs. This result confirms \citet{agostino2019}, which found that X-ray AGNs with weak or no emission lines are not preferentially more distant or in less massive galaxies than AGNs with strong emission lines. 

The reasons for their weak detection appear to be intrinsic, as we do not find hidden bright AGNs but simply weak ones, confirming the assertion from \citet{agostino2023} that the continuum should not affect the detectability of lines because whatever the line flux is, it will simply sit atop of the (higher) continuum with the same strength. Furthermore, our lack of AGN detection in the star-forming galaxy SF-1 confirms the assertion from \citet{agostino2019} that the AGNs powering the X-ray emission in both groups are fundamentally similar in their intrinsic weakness. A similar explanation was previously put forward by \citet{moran1996},  \citet{barger2001}, and \citet{maiolino2003}, and we confirm it is more likely than observational effects such as star formation dilution \citep{moran2002, goulding2009, pons2014, pons2016b}. The dilution scenario in particular was popular in part because it naturally fits in with the mixing sequence picture proposed by \citet{kewley2001b}. However, as shown in \citet{agostino2021}, mixing between SF and AGN emission does not severely affect where an AGN appears on the BPT diagram and is thus unlikely to move AGNs into the SF region of the BPT diagram.

As noted, the possible reasons for the weak emission lines of our X-ray AGNs are intrinsic and include: dust, inefficient UV production, and inefficient UV absorption by the NLR. We do not find evidence for any substantial obscuration of dust, confirming \citet{agostino2023} which found that there is in general not much extra IR emission in AGNs with weak lines. We do not possess any direct evidence supporting or refuting the decreased production of UV photons and it could still play a role. Finally, it is plausible that the absorption rate of ionizing photons by the NLR is inefficient either because of variations in NLR UV absorption efficiency due to a complex geometric distribution of gas \citep{trouille2010}, or the ISM being gas-poor \citep{herpich2018, agostino2023}, or some combination of the two. \citet{herpich2018} looked specifically at how lineless and liny retired galaxies differ so as to determine why some have lines and some do not and suggested that the primary culprit was the amount of gas available to be ionized. While they were focused on a different context, this importance of the availability of gas is conceptually similar to that proposed by \citet{agostino2023} which focused primarily on why some X-ray AGNs had weak [OIII] emission. In the case of the WL-XAGNs in this paper, they tend to have sSFRs below the star-forming main sequence and so it is feasible that they have a lower NLR absorption efficiency to a gas-poor NLR.

\subsection{Completeness of BPT classification}
\label{sec:incompl}
 In this work, we demonstrated that weak-line X-ray AGNs in SDSS show clear AGN signatures when probed with higher spatial resolution spectroscopy, confirming the notion from \citet{agostino2023} that XBONGs or elusive AGNs are not completely lineless.  This demonstrates that current selection methods that require high S/N levels in all four BPT emission lines are going to be biased against weaker AGNs or those in gas-poor galaxies. 
 On the other hand, 2 of the 3 X-ray AGNs in our sample had well-detected [NII]/H$\alpha$ and so it is reasonable to suggest that many of the galaxies in SDSS with AGN-like log([NII]/H$\alpha$)  ($>-0.35$) detected at sufficient S/N and with some detection of [OIII] would also be found to have similar AGN-like signatures with spectroscopy of greater depths or higher spatial resolution than SDSS. This revised selection scheme would successfully include $\sim$95\% of all X-ray AGNs \citep{agostino2023} and therefore increase AGN sample completeness, as suggested by \citet{stasinska2006}.

 \subsection{WHAN diagnostic}

While the BPT diagram remains the most popular tool for separating AGNs and non-AGNs, alternative methods have been employed when data are limited or S/N of the necessary emission lines are low. One such example is the WHAN diagram proposed by
\citet{cidfernandes2011}, which utilizes the equivalent width of H$\alpha$ and the [NII]/H$\alpha$ ratio. By requiring only two emission lines rather than four, one is able to classify more galaxies compared to the BPT diagram, because H$\beta$ is not needed and is typically the weakest emission line. While the BPT diagnostic primarily differentiates between AGNs and non-AGNs, the WHAN diagram provides, on the basis of H$\alpha$ equivalent width, distinctions between strong and weak AGNs as well as two other categories: retired galaxies and passive galaxies. This finer distinction was proposed essentially in an effort to reduce contamination in AGN samples from potential AGN impostors, in particular hot low-mass evolved stars (HOLMES, e.g. \citealt{cidfernandes2011, belfiore2016}).

\begin{figure}
         \begin{center} \includegraphics[width=\columnwidth]{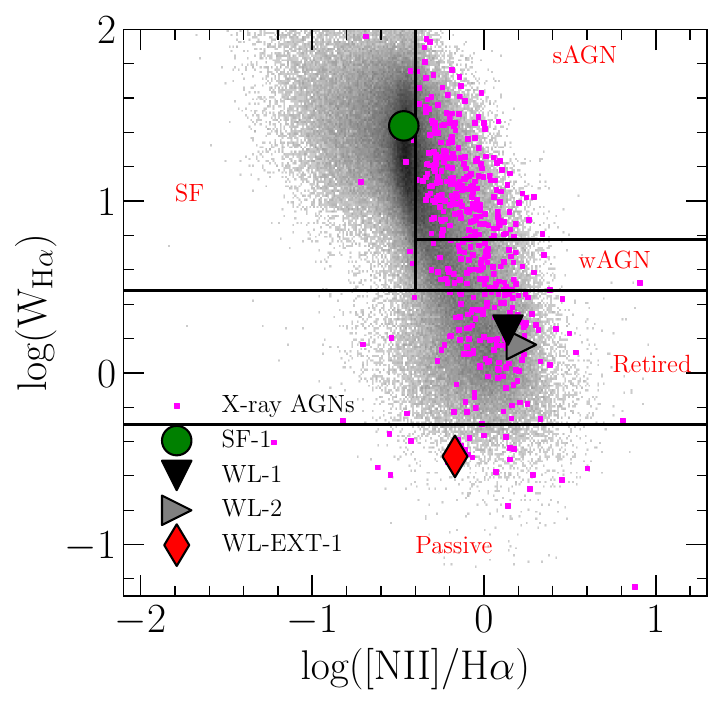}
            \caption{WHAN diagram showing the H$\alpha$ equivalent widths versus the [NII]/H$\alpha$ ratio. Demarcation lines separating the different groups of galaxies come from \citet{cidfernandes2011}. Symbols are the same as in Figure \ref{fig:lxsfr} except the MUSE galaxies possess a black outline so as to better distinguish them from those in the background.
            \label{fig:whan}}
        \end{center}
        \end{figure}

We show the WHAN diagnostic in Figure \ref{fig:whan}, plotting the SDSS-derived values for the X-ray AGNs in the parent sample and highlighting the locations of each of the galaxies we observed with MUSE. The MUSE-derived H$\alpha$ equivalent widths and [NII]/H$\alpha$ values for the galaxies are similar ($<0.1$ difference) to those from SDSS, except for WL-EXT-1 which has its H$\alpha$ equivalent width decrease by 0.4 dex, pushing it further into the passive category. WL-EXT-1's [NII] and H$\alpha$ have S/N between 1 and 2. WL-3 does not have a reliable H$\alpha$ measurement in SDSS or MUSE and so is not included. We opt not to include the other measurements in the plot to prevent unnecessary clutter, but the limited difference in the values in the different apertures suggests that the equivalent widths and the [NII]/H$\alpha$ ratios are not particularly sensitive to the aperture effects. This may be the case for [NII] because the [NII] is produced in partially ionized regions which could extend enough to produce consistent emission line ratios across the physical scales probed here. As for the small differences in H$\alpha$ equivalent width, it is likely a result of the star formation properties not varying substantially enough on the physical scales probed by the 0\farcs 5 and 3\arcsec apertures, as H$\alpha$ equivalent width is primarily a proxy for sSFR.

Despite their AGN-like characteristics (X-ray AGN signature and radially decreasing ionization structure), the WHAN diagnostic does not classify WL-1 or WL-2 as AGNs, but rather as retired galaxies. Indeed, $\sim$1/3 of the X-ray AGNs fall outside of the AGN region of the diagram and within the retired or passive categories, suggesting that using the WHAN diagram for selecting AGNs will result in incomplete samples and that non-AGN samples (for studying retired/passive galaxies) will be contaminated by AGNs. Indeed, it is not clear that the WHAN diagram is able to successfully separate AGNs from HOLMES on the basis of H$\alpha$ equivalent width alone.

\section{Conclusions} 
\label{sec:conclusions}
In this study, we have investigated the spatially-resolved emission line properties of five X-ray AGN candidates (z$\sim$0.1) that have optical emission lines in their SDSS spectroscopy that are atypical for AGNs. With this study, we have been able to address some of the concerns surrounding the
interpretation of SDSS spectroscopy as a result of aperture effects. Our conclusions are the following:

\begin{enumerate}

\item X-ray selected AGNs with undetectable lines in SDSS spectroscopy typically become detectable in deeper VLT-MUSE spectroscopic apertures having the same size (3\arcsec) as SDSS fibers. 

\item There are no cases in which an AGN line is detectable in a very small MUSE aperture (0\farcs5, which is least subject to host contamination) but not in a large, SDSS-like aperture. Furthermore, the line is not more easily detected in a smaller aperture (SNR of the line does not increase). From this we conclude that the continuum swamping or dilution (i.e., aperture effects) is not the ultimate reason why some X-ray AGNs (``optically dull'' AGNs or XBONGs) exhibit weak lines. Rather, the AGN lines are weak in themselves.

\item With the VLT-MUSE spectroscopy, 2 of the 3 AGNs unclassifiable by SDSS are solidly placed in the AGN region of the BPT diagram. In SDSS spectroscopy they had only [NII]/H$\alpha$ detectable, albeit with a value indicative of an AGN.  Our observations suggest that other AGNs that cannot be reliably classified in SDSS by the full BPT diagram (because of weak H$\beta$) but have high [NII]/H$\alpha$ may similarly show definitive AGN signatures with spectroscopy of greater depth. Galaxies with large [NII]/H$\alpha$ should therefore be included in SDSS samples of AGN irrespective of non detections in H$\beta$ or [OIII] (e.g., \citealt{brinchmann04}).

\item The one X-ray AGN in our sample that falls in the star-forming region of the SDSS BPT diagram retains the same position in the BPT diagram even in 0\farcs5 MUSE spectroscopy. The fact that it has not moved into the AGN region demonstrates that apparently ``misclassified'' AGNs are not the result of a process whereby AGN lines of moderate strength become diluted by their host star formation.  Rather, the AGN lines of misclassified AGNs are also very weak to begin with.
\end{enumerate}

\section{Acknowledgements} \label{sec:acknowl}
This research made use of Astropy, a community-developed core Python package for Astronomy \citep{astropy2013} as well as MPDAF \citep{mpdaf}.The construction of GSWLC used in this work was funded through NASA awards NNX12AE06G and 80NSSc20K0440.
CJA would like to thank Natalia Vale Asari for assistance with the \texttt{dobby} software. CJA and SS thank Katherine Rhode and Liese van Zee for thoughtful comments on the material. MB gratefully acknowledges support by the ANID BASAL project FB210003 and from the FONDECYT regular grant 1211000.
We would like to thank the reviewer for providing thoughtful commentary and suggestions which improved the clarity of the manuscript. 


\section{Data Availability}
Reduced VLT-MUSE data products can be made available upon request. Raw data are available through the ESO Science Archive Facility.

\bibliographystyle{aasjournal}
\bibliography{refs}



\appendix


\bsp	
\label{lastpage}
\end{document}